\documentclass[runningheads]{llncs}
\usepackage[T1]{fontenc}
\usepackage{graphicx}
\usepackage{listings}
\usepackage{xcolor}
\usepackage[T1]{fontenc}
\usepackage{courier}  

\lstdefinestyle{llmprompt}{
  basicstyle=\ttfamily\small,
  breaklines=true,
  backgroundcolor=\color{gray!10},
  frame=single,
  columns=fullflexible,
  moredelim=**[is][\bfseries]{@}{@},  
}

\begin{document}
\title{Query Carefully: Detecting the Unanswerables in Text-to-SQL Tasks}
\titlerunning{Query Carefully}

\author{Jasmin Saxer\inst{1}\and
Isabella Maria Aigner\inst{2}\and
Luise Linzmeier\inst{3}\and
Andreas Weiler\inst{1}\and
Kurt Stockinger\inst{1} 
}

\authorrunning{J. Saxer et al.}

\institute{
    Institute of Computer Science, Zurich University of Applied Sciences, Technikumstrasse 9, 8401 Winterthur, Switzerland \and 
    Institute of Medical Virology, University of Zurich, 8057 Zurich, Switzerland \and 
    Department of Gastroenterology and Hepatology, University Hospital Zurich, University of Zurich, 8091 Zurich, Switzerland
    }

\maketitle
\begin{abstract}
Text-to-SQL systems allow non-SQL experts to interact with relational databases using natural language. However, their tendency to generate executable SQL for ambiguous, out-of-scope, or unanswerable queries introduces a hidden risk, as outputs may be misinterpreted as correct. This risk is especially serious in biomedical contexts, where precision is critical. We therefore present Query Carefully, a pipeline that integrates LLM-based SQL generation with explicit detection and handling of unanswerable inputs. Building on the OncoMX component of ScienceBenchmark, we construct OncoMX-NAQ (No-Answer Questions), a set of 80 no-answer questions spanning 8 categories (non-SQL, out-of-schema/domain, and multiple ambiguity types). Our approach employs llama3.3:70b with schema-aware prompts, explicit No-Answer Rules (NAR), and few-shot examples drawn from both answerable and unanswerable questions. We evaluate SQL exact match, result accuracy, and unanswerable-detection accuracy. On the OncoMX dev split, few-shot prompting with answerable examples increases result accuracy, and adding unanswerable examples does not degrade performance. On OncoMX-NAQ, balanced prompting achieves the highest unanswerable-detection accuracy (0.8), with near-perfect results for structurally defined categories (non-SQL, missing columns, out-of-domain) but persistent challenges for missing-value queries (0.5) and column ambiguity (0.3). A lightweight user interface surfaces interim SQL, execution results, and abstentions, supporting transparent and reliable text-to-SQL in biomedical applications.

\end{abstract}
\section{Introduction}
Relational databases are essential tools for managing and querying structured data across numerous domains, including healthcare and scientific research. These systems typically require users to formulate queries using domain-specific programming languages such as SQL (Structured Query Language), which can pose a significant barrier and restrict access to knowledge stored in relational databases to database experts. \\ \\
\textbf{Natural Language to SQL} (text-to-SQL) systems seek to bridge this gap by translating natural language questions into executable SQL queries, enabling users to interact with databases in everyday language without requiring expertise in query languages~\cite{popescu_towards_2003}. Advances in deep learning models and especially Large Language Models (LLM)-based approaches have led to significant improvements in text-to-SQL for simple benchmarking datasets such as Spider, which provide annotated pairs of natural language questions and corresponding SQL queries~\cite{sala_text--sql_nodate,yu_spider_2018}. \\
Despite its potential, text-to-SQL remains a challenging task, especially in sensitive domains such as the healthcare sector, where errors may have serious consequences for patients. Three key challenges include:
\begin{itemize}
    \item Generating accurate and semantically correct queries for complex questions in large domain-specific relational databases.~\cite{singh_survey_2025}
    \item Avoiding the output of misleading or incorrect SQL statements.~\cite{singh_survey_2025}
    \item \textbf{Handling ambiguous, incomplete, or unanswerable user queries}~\cite{zhang_did_2020}.
\end{itemize}

This work focuses on the third challenge: the identification of unanswerable questions. Real users of text-to-SQL systems are typically not database experts. As a result, they may ask questions that are ambiguous, refer to schema elements that do not exist, or lie outside the domain of the database. When confronted with such questions, LLM-based text-to-SQL systems may still attempt to generate SQL queries, resulting in incorrect outputs or misleading results. It is therefore critical that text-to-SQL systems possess the capability to detect unanswerable questions and respond appropriately, for instance, by returning a "not answerable" message or requesting clarification. Detecting unanswerable questions not only helps prevent erroneous results but can also reveal limitations of the system, contributing to the robustness and reliability of text-to-SQL interfaces in real-world applications~\cite{dong_practiq_2024,wang_know_2023,zhang_did_2020,zhao_sphinteract_2024}.

In this paper, we address this challenge by extending an existing text-to-SQL dataset (ScienceBenchmark~\cite{zhang_sciencebenchmark_2023}) with unanswerable questions based on the OncoMX biomedical database containing information about cancer biomarkers. We then propose a pipeline that integrates LLM-based SQL generation with mechanisms to detect and handle unanswerable inputs via an intuitive chat-based interface.

\section{Related Work}
This project is based on the ScienceBenchmark dataset, which contains annotated question-SQL pairs for three complex, domain-specific databases: research policy making (CORDIS), astrophysics (SDSS), and cancer research (OncoMX) \cite{zhang_sciencebenchmark_2023}. We specifically focus on the OncoMX database\footnote{\url{https://www.oncomx.org/}} due to its highly domain-specific biomedical information and terminology. The version of OncoMX utilized in the ScienceBenchmark contains 25 tables with 2 to 14 columns each, for a total of 106 columns and an average of 2,636,771 rows per table. Funded by the U.S. National Institutes of Health (NIH), OncoMX integrates data on cancer biomarkers, differential gene expression in cancer samples, and known cancer mutations from multiple sources, making it an ideal testing database together with question-SQL pairs from the ScienceBenchmark for real-world biomedical text-to-SQL applications.\\ \\
While most existing text-to-SQL research assumes that user queries are valid and answerable within a given schema, few studies have addressed the crucial problem of detecting unanswerable or ambiguous user questions.\\ \\
Zhang et al.~\cite{zhang_did_2020} introduced TriageSQL, a benchmark for classifying user questions into five intention types, including four categories of unanswerable queries (e.g., ambiguous, improper, external knowledge required, and non-SQL). They trained a RoBERTa-based classifier to predict question types based on both the input question and the database schema, highlighting the need for an intention classification step prior to SQL generation. Wang et al.~\cite{wang_know_2023} studied real-world usage of a commercial text-to-SQL product and found that 20\% of user-submitted questions were problematic, revealing the limitations of relying on curated datasets that only include answerable questions. They collected over 3,000 failed user queries and manually grouped them into six categories of unanswerable questions. These categories—focusing on issues such as ambiguity, out-of-scope queries, and schema mismatches—form the basis for the classification framework used in our study, as further described in the methodology section.  Wang et al.~\cite{wang_know_2023} argue that the high frequency of unanswerable queries stems from three key factors: (1) users' unfamiliarity with the database schema, (2) the inherent ambiguity and variability of natural language, and (3) the presence of semantically similar concepts within the database, which can lead to confusion. Wang et al. further focused on two categories—column ambiguity and column unanswerable—and proposed generating unanswerable questions by modifying existing tables rather than creating new unanswerable questions, making the process more controllable but also less realistic.

In the biomedical domain, the EHRSQL 2024 Shared Task addressed the challenge of building reliable text-to-SQL systems for electronic health records (EHR)~\cite{lee_overview_2024}. This task was unique in that it intentionally included unanswerable questions, encouraging systems not only to generate correct SQL queries for answerable inputs but also to abstain from answering when queries were likely unanswerable. Evaluation metrics rewarded systems for producing no SQL for unanswerable questions and penalized both incorrect SQL and attempts to answer when the question was unanswerable. Building on this approach, we applied it to a biomedical dataset (OncoMx) and introduced categorization of unanswerable questions. This allowed us to analyze system performance across different types of unanswerability, with the goal of further enhancing the reliability and safety of text-to-SQL systems in sensitive domains such as healthcare and biomedical research.

\section{Datasets}
\subsection{OncoMX (ScienceBenchmark) }
For the answerable question evaluation, we used the OncoMX dataset provided as part of the ScienceBenchmark suite~\cite{zhang_sciencebenchmark_2023}. Specifically, we evaluated model performance on the development (dev) split of the dataset, which contains a total of 99 natural language questions paired with corresponding SQL queries. In addition to the gold SQL queries, we also retrieved the corresponding results from the OncoMX database. These outputs were used for comparison with model-generated queries and retrieved results.  For few-shot prompting, examples are drawn from the \textit{Seed} dataset of the ScienceBenchmark.

\subsection{OncoMX-NAQ (No-answer Questions)}
To evaluate the performance on finding unanswerable questions, we created a new dataset with 80 questions in 8 categories.

\subsubsection{Categories}
We selected the following unanswerable question categories based on the literature~\cite{wang_know_2023,zhang_did_2020}. 

\paragraph{1. Non-SQL Questions}  
This category includes questions that cannot be answered by any SQL query. Although they may be realistic and relevant to the domain, they are out of scope for structured query generation, typically requiring explanatory or procedural responses instead. For example: "Why does the TP53 gene cause cancer in some patients but not in others?"

\paragraph{2. Out of Schema Questions}  
These questions rely on information that is structurally missing from the schema or not logically accessible within the database. We distinguish two subtypes:
\begin{itemize}
  \item \textbf{Columns Missing:} The question refers to information that is not present in any column of the schema. For example: "What is the 3D protein structure of the EGFR gene product?"
  \item \textbf{Values Missing:} The question depends on specific data values that do not exist in the database, even though the relevant columns may exist.
  For example: "List all genes overexpressed in Martian cancer tissues."
\end{itemize}

\paragraph{3. Out of Domain Questions}  
Questions in this category require knowledge that is external to the database or the application domain (in our case, OncoMX). These questions cannot be resolved without additional outside information. For example: "Which biomarkers are mentioned in the 2023 Nobel Prize research?"

\paragraph{4. Ambiguous Questions}  
This group includes questions that are syntactically valid and appear to be within the scope of the schema, but are underspecified or unclear, resulting in multiple plausible interpretations. Subcategories include:
\begin{itemize}
  \item \textbf{Column Ambiguous:} Multiple columns could reasonably fulfill the request made in the query. For example: "What is the score for EGFR in lung cancer?"
  \item \textbf{Value Ambiguous:} The question references a data value that could correspond to multiple entities or meanings. For example: "Find all genes linked to growth."
  \item \textbf{Contextual Ambiguous:} The intended meaning of the question depends heavily on context, which is not explicitly provided in the query. For example: "What are the genes that cause it?"
  \item \textbf{Operator Ambiguous:} The query implies a comparison or condition, but does not clearly specify which operator (e.g., slightly overexpressed) should be used. For example: "Which biomarkers are more reliable?"
\end{itemize}

\subsubsection{Generation}
To construct the unanswerable question dataset, we generated candidate questions using two distinct prompts with GPT-4o, given the database schema. The generated outputs were manually reviewed and curated by the authors, resulting in a final dataset of 80 unanswerable questions, with 10 representative examples selected for each category. 

The complete set of questions and prompts are available on our GitHub repository\footnote{\label{github}\url{https://github.com/JasminSaxer/QueryCarefully.git}}.

\section{Methodology}
We present \textit{Query Carefully}, a pipeline for querying the OncoMX database with natural language. The method comprises the Pipeline, Prompt Design, and Metrics. The complete code is available on GitHub\footref{github}.

\subsection{Pipeline}

\begin{figure}
    \centering
    \includegraphics[width=1\linewidth]{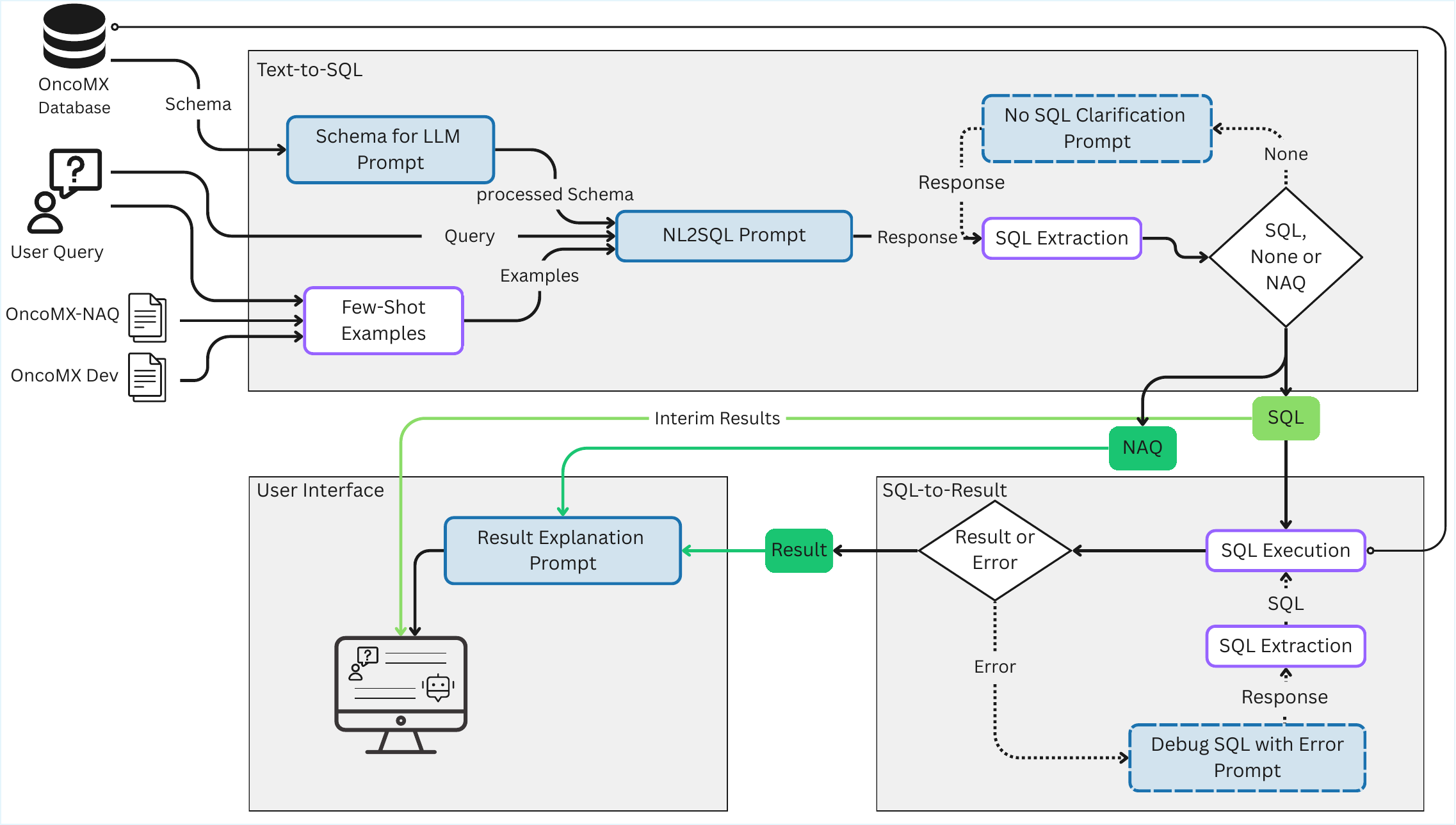}
    \caption{Query Carefully pipeline with user interface. NAQ: No-Answer-Questions.}
    \label{fig:pipeline}
\end{figure}

The full pipeline, including the user interface, is depicted in Figure~\ref{fig:pipeline}.
Our text-to-SQL pipeline employs the large language model \texttt{llama3.3:70b} to translate natural-language questions into executable SQL. For each input question, we construct a task-specific prompt and submit it to the model. The returned text is then parsed to extract a candidate SQL statement. If no SQL is produced and the output does not explicitly indicate that the question is unanswerable, we issue a single standardized re-prompt: \emph{``Please return a SQL query or `unanswerable question' if the question cannot be answered with an SQL query on the database.''}

If a valid SQL query is obtained, it is executed against the target database. In one setting, we include an optional post-processing step for SQL validation. If the query results in an error during execution, we prompt the LLM again with the error message:  
\textit{``Please correct the SQL query based on the following error message: <error>''}.  
This correction loop is allowed up to a maximum of three retries.

\subsection{Prompt Design}
To ensure high-quality SQL generation in our text-to-SQL setting, we carefully designed a prompt template tailored to the OncoMX relational database schema. The base prompt included a system message instructing the LLM to act as an expert in natural language to SQL translation and to generate syntactically correct PostgreSQL queries. The model was instructed to return \textbf{only} the SQL query, without additional explanation or commentary. In addition, the prompt included the reformatted and human-readable database schema. 

\paragraph{Schema Representation}  
The database schema was reformatted to be more LLM-friendly using GPT-4o, ensuring that table and column names are readable and semantically meaningful. The full schema was presented to the model in a structured and human-readable format, listing each table along with its columns, data types, foreign key relationships, and column-level comments to preserve semantic context. 

\paragraph{Handling Unanswerable Questions}  
To ensure that the model does not attempt to generate SQL queries for questions that cannot be reliably answered from the database, we included explicit instructions in the system prompt, taking a similar approach to \cite{jo_lg_2024}. These rules were designed to guide the model to return \texttt{"unanswerable question"} when the questions shouldn't be answered.

\paragraph{User Prompt}  
The user prompt consisted of a natural language query prefixed with a short instruction. 
For example:
\begin{quote}
\texttt{\# Return the SQL for the following Question}\\
\texttt{[Q]: Show me all disease mutations with ref\_aa E}\\
\texttt{[SQL]:}
\end{quote}

\paragraph{Few-Shot Examples.}  
To assess the effect of in-context learning, we evaluated four prompting configurations: zero-shot, 1-shot, 3-shot, and 5-shot. In the few-shot settings, the prompt was augmented with examples of both answerable and unanswerable questions. Answerable examples included correctly grounded SQL statements, while unanswerable examples featured natural language questions followed by the response \texttt{unanswerable question}.

To select relevant few-shot examples dynamically, we used the \textit{Alibaba-NLP /gte-Qwen2-1.5B-instruct} model to embed the natural language questions. The most similar answerable questions were retrieved from the seed OncoMX ScienceBenchmark dataset. The most similar unanswerable questions were retrieved from our new dataset, removing the question to be answered. Cosine similarity was used as the measure of similarity.

A full prompt example is shown in GitHub.

\subsection{Metrics}
To comprehensively evaluate our text-to-SQL system, we report performance across three distinct metrics, each capturing a different aspect of system behavior: syntactic correctness, semantic accuracy, and robustness to unanswerable queries. Each metric is computed over the answerable or unanswerable questions dataset and reported as an accuracy score: the number of correct predictions divided by the total number of evaluated queries. 

\paragraph{SQL Exact Match Accuracy} 
This metric checks whether the predicted SQL string exactly matches the gold (reference) SQL query after normalization (i.e., converting to lowercase and removing extra whitespace). It is a strict comparison that does not tolerate semantically equivalent but syntactically different queries.

\paragraph{Result Accuracy}  
This metric evaluates the semantic correctness of a predicted SQL query by comparing its execution result with that of the gold (reference) query. It supports multiple levels of comparison to account for variations in formatting or minor structural differences:

\begin{itemize}
    \item \textit{Exact Match:} The result tables are identical in both content and structure, including row order and column names.
    \item \textit{Soft Correct:} The predicted and gold results contain the same data, allowing for differences in row order, column names, and the presence of identifier columns (e.g., \texttt{id}). The soft correct score is inspired by \cite{nooralahzadeh_statbotswiss_2024}.
    \item \textit{DB Error:} The predicted SQL query could not be executed successfully due to a database error (e.g., syntax error or invalid reference).
\end{itemize}

\paragraph{Unanswerable Question Detection}  
This metric evaluates the system's ability to correctly identify questions that cannot be answered. It checks whether the model returns the indicator \texttt{"unanswerable question"} for queries labeled as unanswerable in the gold data. Additionally, we also verify that the model does not return \texttt{"unanswerable question"} for answerable queries.

\section{Results}
\begin{figure}[h]
    \centering
    \includegraphics[width=0.7\textwidth]{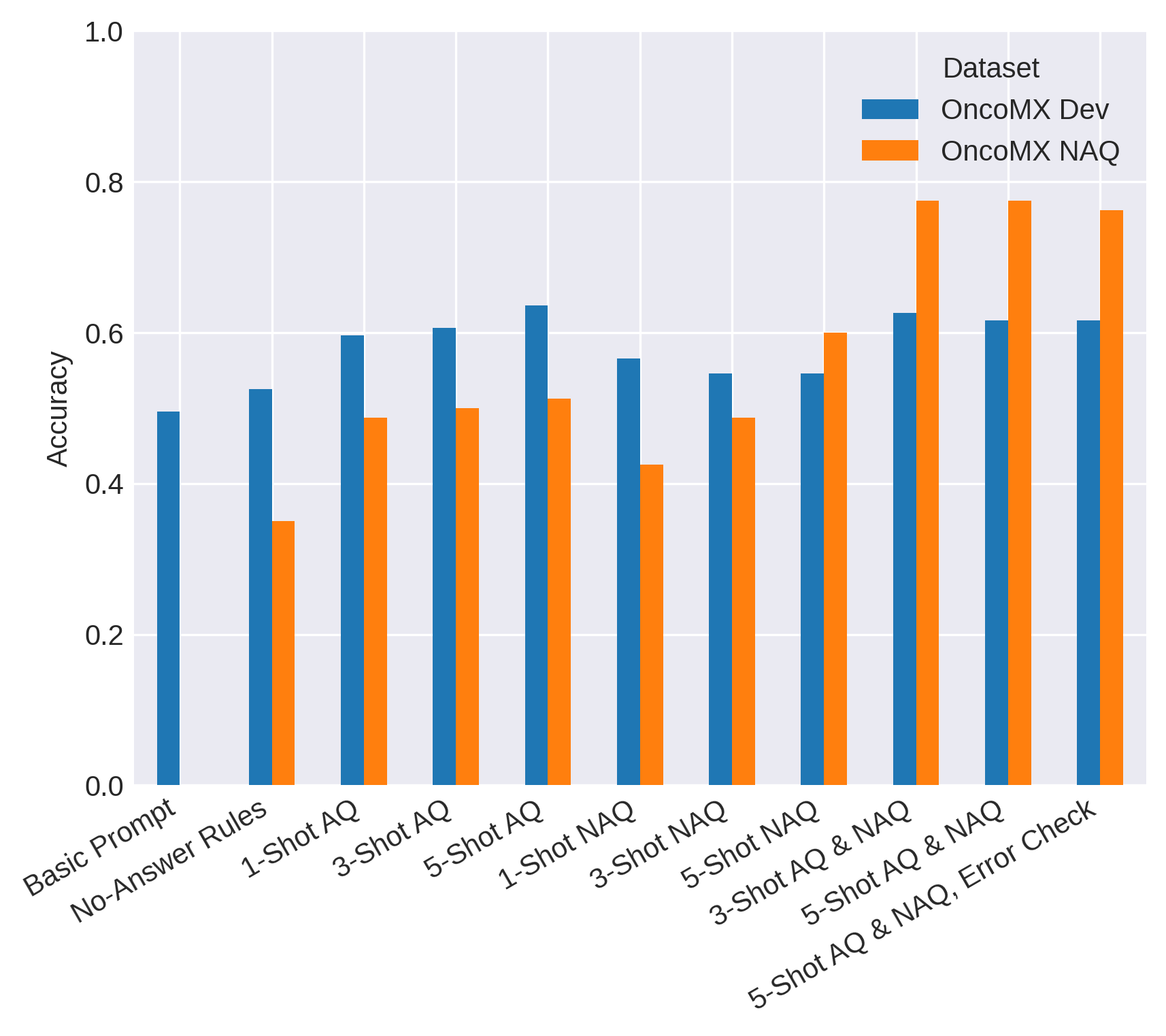}
    \caption{Accuracy of different prompts on OncoMX Dev and OncoMX-NAQ dataset. The soft correct accuracy of OncoMX Dev is shown.}
    \label{fig:datasets}
\end{figure}

We evaluated prompt designs on the OncoMX Dev dataset and on OncoMX-NAQ, our extension of OncoMX Dev with 10 additional unanswerable questions per category (Fig \ref{fig:datasets}). On OncoMX Dev, adding the "No-Answer Rules" (NAR) alone had no effect on the accuracy (accuracy ~0.5). Including answerable examples modestly improved accuracy to 0.6, with the 5-shot prompt performing best. Adding both answerable and unanswerable examples produced similar accuracies, showing that positive examples enhance answerability detection and unanswerable examples do not impact performance on datasets without unanswerable questions.
On our OncoMX-NAQ dataset, the NAR alone resulted in an accuracy of 0.37. Adding answerable examples improved performance regardless of in-context learning, though it stayed below OncoMX dev levels. Increasing unanswerable examples further boosted accuracy, surpassing OncoMX dev at 5-shot. The highest accuracy (close to 0.8) was achieved with both answerable and unanswerable examples, highlighting the value of balanced prompting for datasets containing unanswerable questions.

\subsection{OncoMX dev}
\begin{figure}[h]
    \centering
    \includegraphics[width=1\textwidth]{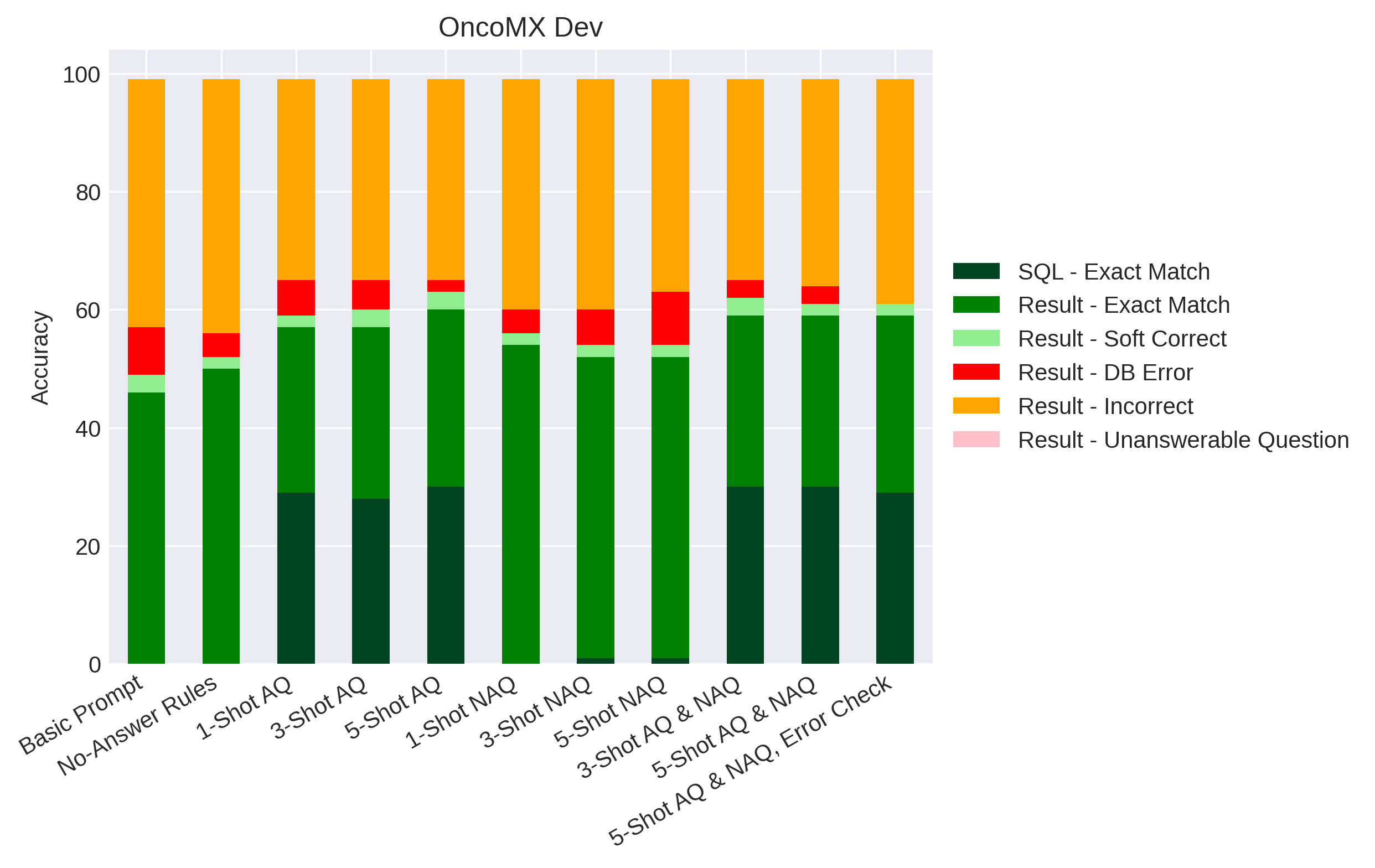}
    \caption{Different measurements of ScienceBenchmark (OncoMX dev) per prompt type. AQ: Answerable Questions, NAQ: No-Answer Questions.}
    \label{fig:diffacc}
\end{figure}

Figure~\ref{fig:diffacc} compares accuracy metrics on the OncoMX dev set across prompt variants. Exact SQL matches are observed only when using in-context learning (ICL) with answerable examples from the OncoMX Seed split, indicating that a strict SQL exact match is an insufficient standalone metric for text-to-SQL evaluation. Across all prompts, the \textit{soft-correct} category occurs only rarely; in most cases, the system either reproduces the gold result set exactly or diverges substantially. Introducing the error-checking loop removes database execution errors but increases incorrect results, suggesting that syntactic validation alone does not guarantee factual correctness and that many queries remain intrinsically challenging for the model.

\subsection{Accuracy by No-Answer Question Type}
\begin{figure}[h]
    \centering
    \includegraphics[width=1\textwidth]{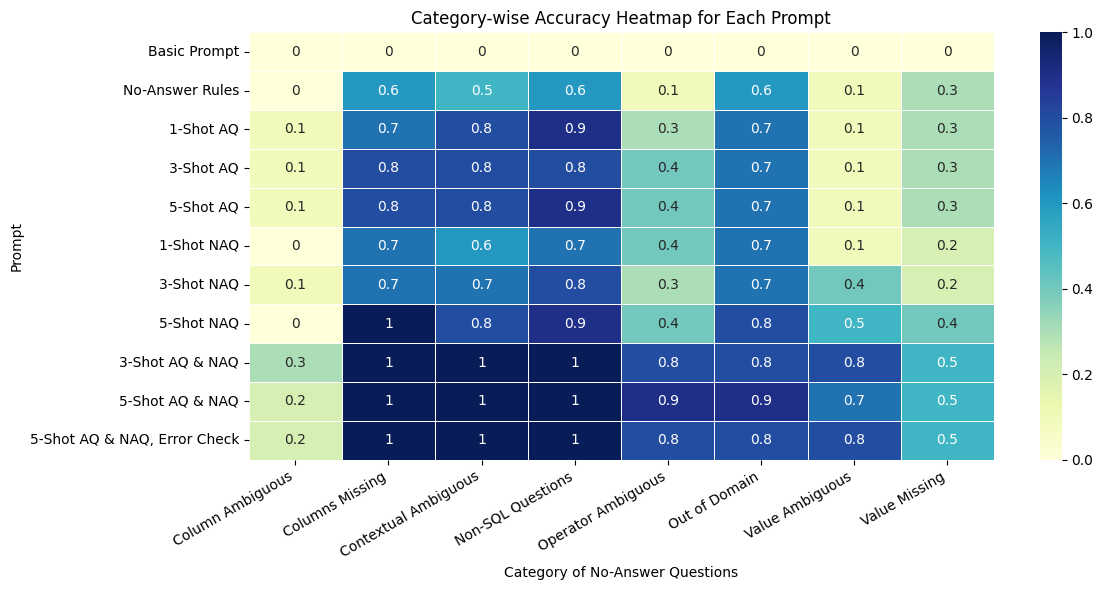}
    \caption{Accuracy of detection of unanswerable questions divided by category and prompt. AQ: Answerable Questions, NAQ: No-Answer Questions. }
    \label{fig:category_wise}
\end{figure}

We now analyze how unanswerable questions are detected. To further dissect the effect of the NAR on different categories of unanswerable questions, we performed a more detailed unanswerable question category-wise analysis (Fig \ref{fig:category_wise}). This revealed differences in the detectability of unanswerable questions across types:

\begin{itemize}
    \item Structurally defined categories such as \textit{Non-SQL}, \textit{Columns Missing}, and \textit{Out of Domain} showed moderate baseline accuracy with the NAR alone (0.6–0.65). Adding either answerable or unanswerable examples improved accuracy substantially, with prompts containing both types reaching near-perfect detection for these categories.
    \item In the \textit{Values Missing }category, baseline accuracy was low (0.3) and improved only when both answerable and unanswerable examples were included, reaching a maximum of 0.5. This reflects the difficulty of detecting missing values from schema information alone, as missing columns can be directly inferred from schema details, while missing values cannot.
    \item Ambiguous categories (\textit{Column Ambiguous}, \textit{Value Ambiguous}, \textit{Contextual Ambiguous}, \textit{Operator Ambiguous}) showed more varied patterns. Notably, \textit{Column Ambiguous} questions were poorly classified (max accuracy 0.3), as text-to-SQL often generated SQL referencing one fitting column without recognizing ambiguity. By contrast, \textit{Value Ambiguous} questions achieved higher accuracy, especially when unanswerable examples were included, indicating the model better handles ambiguity over values than columns.
    \item For \textit{Contextual Ambiguous} questions, about half were detected by the NAR alone, with full accuracy only achieved when both answerable and unanswerable examples were provided. Similarly, \textit{Operator Ambiguous }questions required both example types to reach high accuracy (0.9). One misclassified question was ultimately deemed answerable due to the presence of a relevant database column.
\end{itemize} 

Interestingly, including answerable examples alone improved unanswerable question detection in nearly all categories more than including unanswerable examples alone (average accuracy 0.49 vs. 0.43, compared to 0.35 with NAR only). However, combining both example types consistently yielded the highest accuracy in all categories.

\subsection{Text-to-SQL User Interface}
\begin{figure}[h!]
    \centering
    \includegraphics[width=1\linewidth]{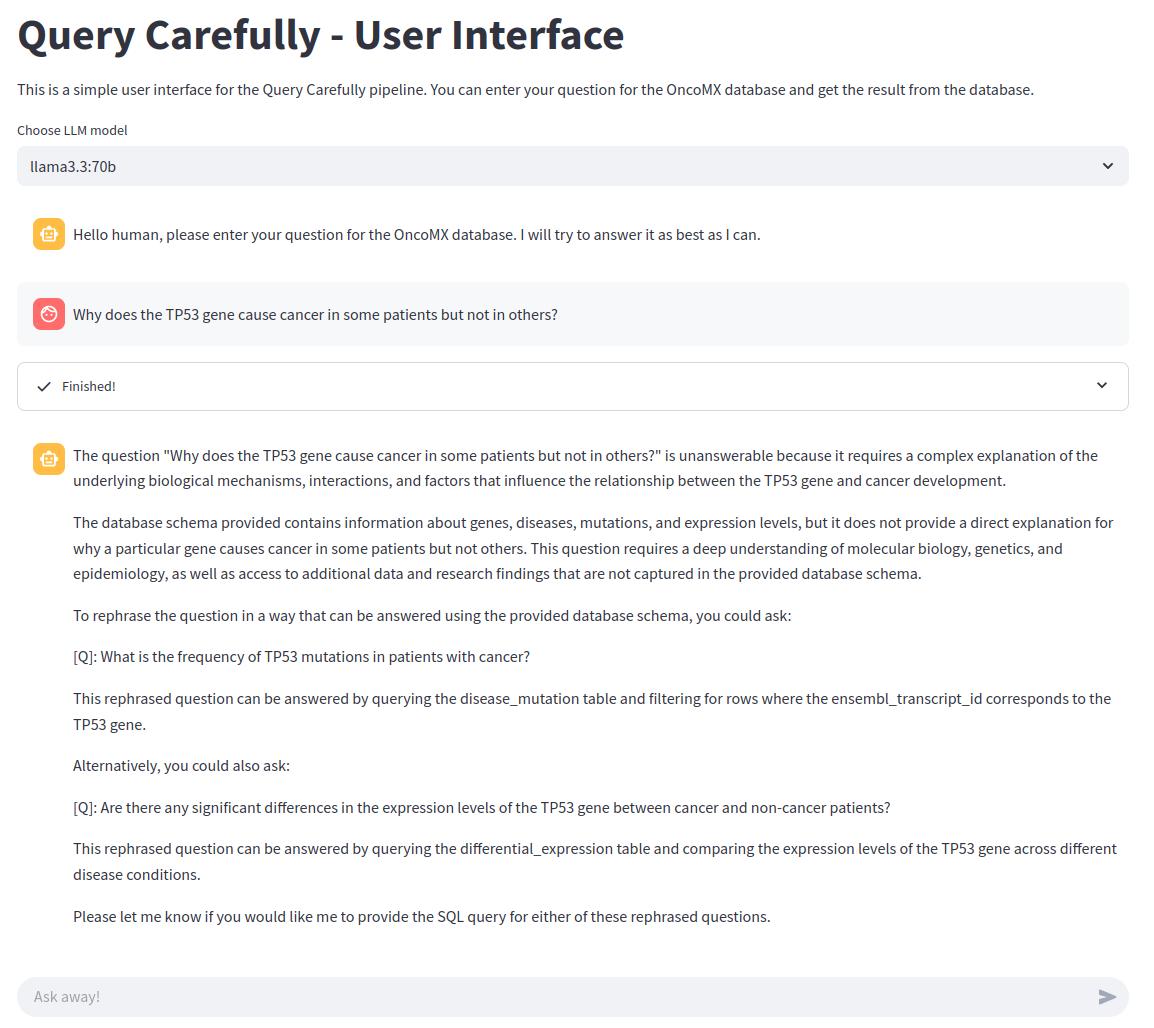}
    \caption{Example of an unanswerable question on the user interface.}
    \label{fig:ui}
\end{figure}

We provide a lightweight web interface for interacting with the \textit{Query Carefully} pipeline (see Figure~\ref{fig:ui}). Users select the large language model from a drop-down menu (default: \texttt{llama3.3:70b}), enter a natural language question about the OncoMX database, and receive the generated SQL, the corresponding result table, and a short answer to the question. The interface surfaces intermediate steps, including the SQL preview, execution status, and any error messages. By making the SQL and outputs explicit, the UI supports transparency, easy verification by domain experts, and rapid iteration on query phrasing. When the model detects an unanswerable question, it returns an explanation why it can't answer the question and gives methods for improvement. 

\section{Conclusion}
Text-to-SQL systems should account for user expertise and potential unanswerable queries, a limitation in current benchmark datasets, particularly in domains such as medicine where non-SQL experts may misinterpret results. Existing datasets like PRACTIQ \cite{dong_practiq_2024} classify queries based on SQL statements into categories such as ambiguous or nonexistent columns, filters, and joins. In contrast, our work extends this approach by also including non-SQL and out-of-domain questions, capturing a broader range of real-world user queries. Conversational approaches, such as PRACTIQ and Sphinteract \cite{zhao_sphinteract_2024}, enhance user interaction by providing explanations for unanswerable queries. Nevertheless, effective use requires that users either identify the relevant columns themselves or are provided with explicit schema annotations to guide them to find the matching query. Methods like "Disambiguate First, Parse Later" (Saparina \& Lapata, 2025) and "Is Long Context All You Need" (Chung, 2025) show that generating multiple interpretations or leveraging extended context windows can help resolve ambiguity, albeit with higher computational cost. Our results align with prior findings that example-based prompting improves text-to-SQL question classification \cite{zhao_sphinteract_2024,dong_practiq_2024}, but uniquely demonstrate that answerable examples can contribute more to detecting unanswerable queries than negative examples, highlighting the importance of balanced example inclusion. 
\\ \\
Limitations of our study include challenges in detecting specific categories of unanswerable queries, particularly those involving missing values or column and value ambiguities. Furthermore, identifying unanswerable queries represents only an initial stage. Incorporating mechanisms to propose potential resolutions, as suggested in PRACTIQ or Sphinteract could further enhance the robustness of text-to-SQL systems.

\begin{credits}

\subsubsection{\discintname}
The authors have no competing interests to declare that are relevant to the content of this article.

\end{credits}
\bibliographystyle{splncs04}
\bibliography{references}

\end{document}